# Effect of coupling on scheme of hysteresis jumps in current-voltage characteristics of intrinsic Josephson junctions in high-$T_c$ superconductors

Yu.M.Shukrinov, and F.Mahfouzi

*Abstract*—We report the numerical calculations of the current-voltage characteristics of intrinsic Josephson junctions in high-$T_c$ superconductors. The charging effect at superconducting layers is taken into account. A set of equations is used to study the non-linear dynamics of the system. In framework of capacitively coupled Josephson junctions model we obtain the total number of branches using fixed initial conditions for phases and their derivatives. The influence of the coupling constant α on the current-voltage characteristics at fixed parameter β ($\beta^2 = 1/\beta_c$, where $\beta_c$ is McCumber parameter) and the influence of α on β-dependence of the current-voltage characteristics are investigated. We obtain the α-dependence of the branch's slopes and branch's endpoints. The obtained results show new features of the coupling effect on the scheme of hysteresis jumps in current-voltage characteristics of intrinsic Josephson junctions.

## I. Introduction

The phase dynamics in the intrinsic Josephson junctions (IJJ) has attracted a great interest because of rich and interesting physics from one side and perspective of applications from the other one. Different type of couplings between junctions, like inductive coupling in the presence of magnetic field [1], capacitive [2]-[3], charge-imbalance [4] and phonon [5] couplings determine a variety of current-voltage characteristics (IVC) observed in high temperature superconductors ( HTSC). In [6] has been stressed that capacitive coupling takes various values in HTSC and layered organic superconductors, that is, the capacitive coupling is tunable in these systems. Based on this fact a study for the dynamics of the CCJJ model, focusing on the dependence of phase dynamics on the strength of the capacitive coupling constant has been presented in this paper.

Yu.M.Shukrinov is with the BLTP, JINR, Dubna, Moscow Region, 141980, Russia and Physical Technical Institute, Dushanbe, 734063, Tajikistan.
F.Mahfouzi is with the Institute for Advanced Studies in Basic Sciences, P.O.Box 45195-1159, Zanjan, Iran.
This work was supported in part by the INTAS under Grant No. 01-0617.

## II. Model and Numerical Results

In the CCJJ model the dynamics of the gauge-invariant phase difference $\varphi_l$ between superconducting layers $l$ and $l+1$ is described by the equation:

$$\ddot{\varphi}_l = \frac{I}{I_c} - \beta\dot{\varphi}_l - \sin\varphi_l - \alpha(2\sin\varphi_l - \sin\varphi_{l+1} - \sin\varphi_{l+1}) \quad (1)$$

where $I$ and $I_c$ are the external dc current and the Josephson critical current , respectively.

We proposed a fixed initial conditions method (FIC-method), which is based on determination of the initial conditions using the values of branch's slopes. By this method we simulate the IVC of IJJ under restriction that patterns of distribution of phase rotating junctions are symmetric [7]. For the case of 11 junctions at α=1, β=0.2, γ=0.5 we obtain the complete branch structure consisting of 45 branches with a different slopes.

The influence of the coupling parameter α on the IVC of a stack of IJJ is demonstrated in Fig.1, where the IVC calculated at fixed initial conditions are shown at α=0.1, 0.5 and 1. The main features of this influence which we are concentrating on in this paper are the change of the slopes and the endpoints of branches. As seen in Fig.1, the resistive branches shift towards the higher voltage side (towards the outermost branch ) [8] and their's endpoints are increasing with increase in α.

Fig.2 shows the α-dependence of the branch's slopes for some branches. The slope of the outermost branch (the all junctions are in the rotating (R-state)) does not depend on the value of the coupling constant. As we expected the slopes of the branches are getting close to the slope of the outermost branch, but this approaching is decreased with increase in α. The α-dependence of the branch's slopes demonstrates the change of branch's order with increase in the α.

Some curves are saturated at definite values of slope n. The different behaviour depends on whether the state includes the first and last junction (boundary conditions), whether two or more junctions in oscillating state (O-states) are neighbors or they separated by junctions in R-state.

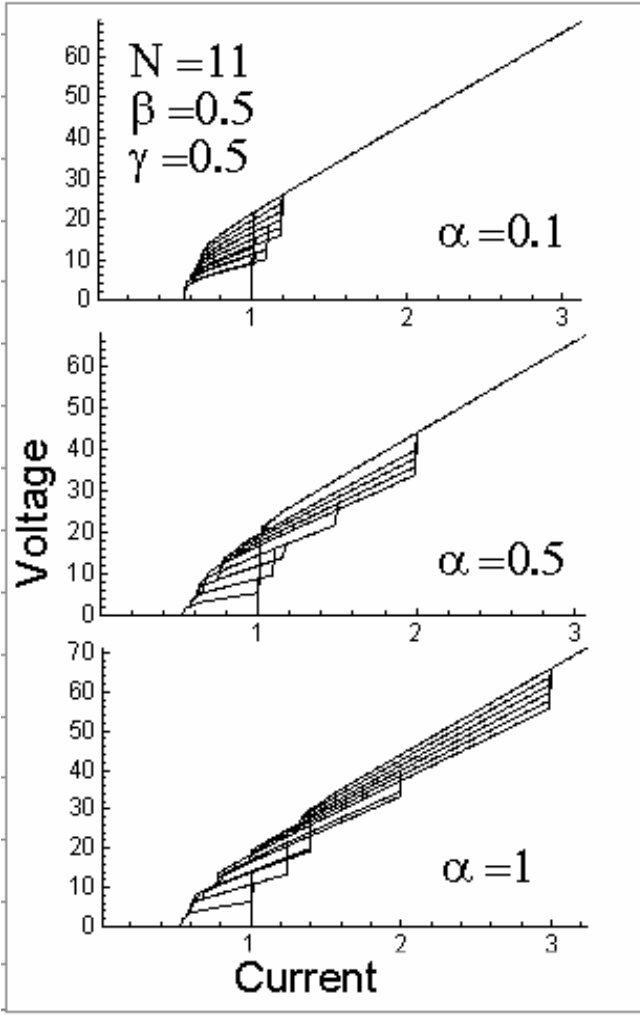

Fig. 1. The change of the IVC with increase in coupling parameter $\alpha$.

Using the equations of CCJJ model [7] we obtain the analytical expression for the α-dependence of the slope n, taking into account the distribution of R- and O-junctions in the stack. For example,

$$n = \begin{cases} N - \dfrac{2+\alpha}{1+1.5\alpha} & for\ state\ O(1,11) \\ N - \dfrac{3+10\alpha}{1+4\alpha+2\alpha^2} & for\ state\ O(5,6,7) \end{cases} \quad (2)$$

The slope for the branch O(5,6,7) (junctions 5,6,7 are in O-state) limits to the slope of the outermost branch at $\alpha \to \infty$. As we mentioned before, the order of the branches in IVC is changed with increase in α. For example, the position of the branches 31 and 21 have changed. The coupling between junctions breaks the equidistance of the branch structure and it happens at enough small value of α.

In general, each junction in O-state in the stack has its own α-dependence of the phase difference and that one, which has the most strong dependence, determines the α-dependence of the branch's endpoints. From the analysis of equation (1) and resistively shunted junction equation we find that, for example, for state O(5,6,7) the phase difference in junction 6 determines the endpoint. Mostly the α-dependences of the endpoints are monotonic, but in some cases the strongest α-dependence of $\sin\varphi_l$ is transformed from one junction to another one with increase in α. It leads to a broken dependence. For example, for branch 28 (state O(1,5,6,7,11)) α-dependence is determined by junction 6 at small α, but by junctions 1 and 11 at big α. The analytical dependence in this case has a form

$$I^{end}(\alpha) = \begin{cases} \dfrac{1+4\alpha+2\alpha^2}{1+4\alpha} & at\ \alpha < 1+\sqrt{2} \\ \dfrac{1+1.5\alpha}{1+0.5\alpha} & at\ \alpha < 1+\sqrt{2} \end{cases} \quad (3)$$

Results of simulation of the IVC at different values of α show the different $\beta$-dependence of the return current for the first and last branches which are different from such dependences for one junction.

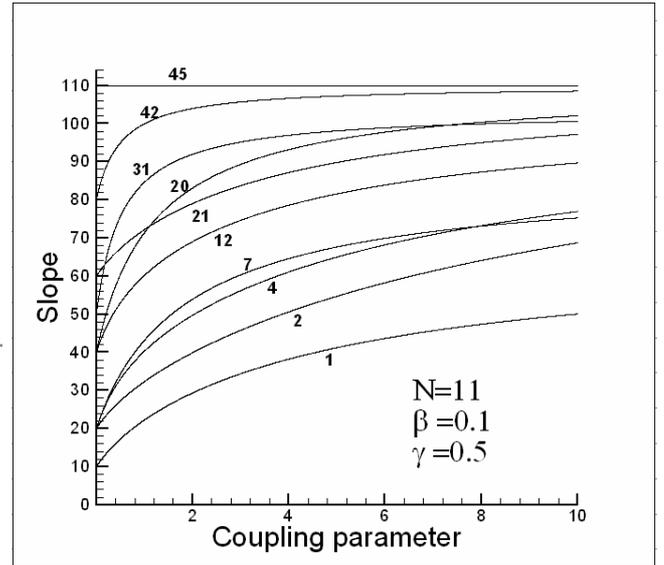

Fig. 2. The $\alpha$-dependence of the branch's slopes.

We consider that comparison the obtained results with experimental data allow to determine the limits of CCJJ model and show the way to improve it.

REFERENCES

[1] Kleiner R, Steinmeyer F, Kunkel G, and Muller P 1992 Phys. Rev. Lett. **68** 2394
[2] Oya G, Aoyama N, Irie A, Kishida S, and Tokutaka H 1992 Jpn. J. Appl. Phys. **31** L829
[3] Bulaevskii L N, Dominguez D, Maley M, Bishop A, and Ivlev B 1995 Phys. Rev. B **53** 14601
[4] Koyama T, and Tachiki M 1996 Phys. Rev. B **54** 16183
[5] Machida M, Koyama T, and Tachiki M 1998 Physica **C300** 55
[6] Machida M, Koyama T, and Tachiki M 1999 Phys.Rev.Lett. **83** 4816
[7] Matsumoto H, Sakamoto S, Wajima F, Koyama T, Machida M 1999 Phys. Rev. B **60** 3666
[8] Machida M, Koyama T 2004 Phys. Rev. B **70** 024523